\documentclass[12pt]{article}
\usepackage[letterpaper,margin=0.6in]{geometry}
\usepackage{bm}
\usepackage{authblk}
\usepackage{amsmath}
\usepackage{graphicx}
\usepackage[section]{placeins}
\usepackage{xcolor}
\usepackage{wrapfig}
\usepackage{wrapfig,lipsum,booktabs}
\usepackage{tabularx}
\usepackage{float}
\usepackage{newtxtext,newtxmath}
\makeatletter
\let\saved@includegraphics\includegraphics
\AtBeginDocument{\let\includegraphics\saved@includegraphics}
\renewenvironment*{figure}{\@float{figure}}{\end@float}
\makeatother
\usepackage[labelfont=bf]{caption}
\usepackage{scicite}
\usepackage{soul}

\usepackage{hyperref}
\hypersetup{colorlinks=true, linkcolor=blue, citecolor=blue, urlcolor=blue,}

\def\scititle{
Colossal enhancement of spin transmission through magnon confinement in an antiferromagnet
}
\title{\bfseries \boldmath \scititle}
\author[1,2$\dagger$,$\ast$]{Sajid Husain}          
\author[3,$\dagger$]{Maya Ramesh}
\author[4]{Xinyan Li}
\author[5]{Sergei Prokhorenko}
\author[4]{Shashank Kumar Ojha}
\author[6]{Aiden Ross}
\author[7,8]{Koushik Das}
\author[9]{Boyang Zhao}
\author[10]{Hyeon Woo Park}
\author[1]{Peter Meisenheimer}
\author[5]{Yousra Nahas}
\author[11]{Lucas Caretta}
\author[2,4,12]{Lane W. Martin}
\author[10]{Se Kwon Kim}
\author[13]{Zhi Yao}
\author[9,14]{Haidan Wen}
\author[7,15]{Sayeef Salahuddin}
\author[6]{Long-Qing Chen}
\author[16]{Yimo Han}
\author[17,18]{Rogério de Sousa}
\author[5,19]{Laurent Bellaiche}
\author[20]{Manuel Bibes}
\author[3,21,22]{Darrell G. Schlom}
\author[1,2,4,15,23$\ast$]{Ramamoorthy Ramesh}

\affil[1]{Department of Materials Science and Engineering, University of California, Berkeley, CA 94720, USA}
\affil[2]{Materials Science and NanoEngineering, Rice University, Houston, Texas, USA}
\affil[3]{Department of Materials Science and Engineering, Cornell University, Ithaca, NY 14853, USA}
\affil[4]{Rice Advanced Materials Institute, Rice University, Houston, TX, 77005, USA}
\affil[5]{Smart Ferroic Materials Center, Physics Department and Institute for Nanoscience and Engineering, University of Arkansas, Fayetteville, Arkansas, USA}
\affil[6]{Department of Materials Science and Engineering and Materials Research Institute, Pennsylvania State University, University Park, Pennsylvania 16802, USA}
\affil[7]{Department of Electrical Engineering and Computer Sciences, University of California, Berkeley, CA 94720, USA}
\affil[8]{Department of Chemistry, University of California, Berkeley, CA 94720, USA}
\affil[9]{Materials Science Division, Argonne National Laboratory, Lemont, IL 60439, USA.}
\affil[10]{Department of Physics, Korea Advanced Institute of Science and Technology, Daejeon, Korea}
\affil[11]{School of Engineering, Brown University, Providence, RI, 77005, USA}
\affil[12]{Departments of Chemistry and Physics and Astronomy, Rice University, Houston, TX, 77005, USA}
\affil[13]{Applied Mathematics and Computational Research Division, Lawrence Berkeley National Laboratory, CA 94720, USA}
\affil[14]{Advanced Photon Source, Argonne National Laboratory, Lemont, IL 60439, USA}
\affil[15]{Materials Science Division, Lawrence Berkeley National Laboratory, Berkeley, CA, 94720, USA}
\affil[16]{Rice Advanced Materials Institute, Smalley-Curl Institute, and Ken Kennedy Institute, Rice University, Houston, Texas 77005, USA}
\affil[17]{Department of Physics and Astronomy, University of Victoria, Victoria, British Columbia, Canada V8W 2Y2}
\affil[18]{Centre for Advanced Materials and Related Technology, University of Victoria, Victoria, British Columbia, Canada V8W 2Y2}
\affil[19]{Department of Materials Science and Engineering, Tel Aviv University, Ramat Aviv, Tel Aviv 6997801, Israel}
\affil[20]{Laboratoire Albert Fert, CNRS, Thales, Université Paris-Saclay, 91767 Palaiseau, France}
\affil[21]{Kavli Institute for Nanoscale Science, Cornell University, Ithaca, NY 14853, USA.}
\affil[22]{Leibniz-Institut für Kristallzüchtung, 12489 Berlin, Germany.}
\affil[23]{Department of Physics, University of California, Berkeley, CA, 94720, USA}

\graphicspath{{figures/}}

\date{}                   
\begin{document}
	\maketitle
	$^\ast$  E-mail: shusain@berkeley.edu, rramesh@berkeley.edu
        $^\dagger$ Authors contributed equally.
\newpage	
	\vspace{10pt}
\linespread{1.2}	
	\begin{abstract}
		\textbf{Since Felix Bloch's introduction of the concept of spin waves in 1930, magnons (the quanta of spin waves) have been extensively studied in a range of materials for spintronics, particularly for non-volatile logic-in-memory devices. Controlling magnons in conventional antiferromagnets and harnessing them in practical applications, however, remains a challenge. In this letter, we demonstrate highly efficient magnon transport in an LaFeO$_3$/BiFeO$_3$/LaFeO$_3$ all-antiferromagnetic system which can be controlled electrically, making it highly desirable for energy-efficient computation. Leveraging spin-orbit-driven spin-charge transduction, we demonstrate that this material architecture permits magnon confinement in ultrathin antiferromagnets, enhancing the output voltage generated by magnon transport by several orders of magnitude, which provides a pathway to enable magnetoelectric memory and logic functionalities. Additionally, its non-volatility enables ultralow-power logic-in-memory processing, where magnonic devices can be efficiently reconfigured via electrically controlled magnon spin currents within magnetoelectric channels.}
	\end{abstract}
  
More than 50 years after the famous comment by Louis Néel in his Nobel lecture that antiferromagnets are ``interesting but useless", an exciting era of antiferromagnetic spintronics is rapidly emerging\cite{baltz2018antiferromagnetic}. Although it is true that controlling antiferromagnets with tools commonly used for ferromagnets, such as a magnetic field, is theoretically possible, but requires impractically large magnetic fields. However, the emergence of spin-transport physics has completely transformed this perspective by enabling electrical manipulation\cite{jungwirth2016antiferromagnetic}. Modern control of antiferromagnets takes advantage of a variety of approaches such as electromagnetic radiation\cite{nvemec2018antiferromagnetic}, relativistic current-induced fields in broken inversion-symmetry antiferromagnets through the inverse spin-galvanic effect/Rashba-Edelstein effect\cite{wadley2016electrical,vzelezny2014relativistic,bhattacharjee2014prediction}, as well as the spin-Hall effect\cite{duttagupta2020spin}. The classical model systems for such demonstration are canted antiferromagnets such as $\alpha$-Fe$_2$O$_3$\cite{lebrun2018tunable}, or ferrimagnetic insulators such as yttrium-iron garnet\cite{wei2022giant}, and many more\cite{han2023coherent} wherein studies have led to some striking discoveries\cite{parsonnet2022nonvolatile}, such as a strong enhancement in the inverse spin-Hall voltage with material and device dimensions\cite{wei2022giant,huang2024manipulating}. Antiferromagnets present several advantages as compared to ferromagnets such as a larger magnon-group velocity\cite{hamdi2023spin}, a significantly higher antiferromagnetic resonance frequency ranging from several hundred GHz to THz (which is compatible with the frequency ranges for future telecommunications such as 6G and beyond), and insensitivity to external magnetic fields (avoiding cross-talk between the information bits). Most importantly, a wide range of antiferromagnets are insulators, and thus there is a strong potential to reduce Joule energy losses during transmission which improves the overall energy efficiency. 

From a macroscopic perspective, energy efficiency in computing has become an increasingly pervasive global challenge, triggering numerous parallel pathways aimed at addressing it\cite{AlbertRMP,salahuddin2018era}. Among those are intriguing ideas that involve creating an in-memory compute element using the bistable ferroelectric state of a multiferroic for non-volatile storage and the spin component to carry out logic operations in a so-called magnetoelectric spin-orbit (MESO) logic\cite{manipatruni2019scalable}. A key recognition was that the operating voltage translates to the energy consumed in the logic/memory element - a requirement that drives the need to achieve sub-100 mV operation to reach the desired attoJoule/operation computing element scale. %Such dramatic advances are required to achieve the power reductions necessary to mitigate its growth\cite{ramesh2024roadmap}. 
Such devices work with a voltage input which is converted by the ME multiferroic into a spin signal through a ferromagnet which is used to carry out the logic operations. Reading out the spin state is accomplished through spin-charge conversion, $e.g.$, by the inverse-spin-Hall effect (ISHE)\cite{vaz2024voltage}. The original MESO concept\cite{manipatruni2019scalable} used a ferromagnetic layer in contact with the multiferroic to help read the magnetic state. However, recent works\cite{chai2024voltage,huang2024manipulating} have demonstrated that the antiferromagnetic state in bismuth ferrite `BiFeO$_3$' (BFO) can be manipulated with an electric field (for writing)\cite{cherifi2014electric,chu2008electric} and sensed by the ISHE through a SO metal (for reading)\cite{parsonnet2022nonvolatile}.
The requirement of obtaining $>$ 100 mV output through the spin-to-charge conversion ($V_{ISHE}$)
has been identified as a materials physics ``grand challenge". Indeed, until recently, the magnitude of the $V_{ISHE}$ (with Pt) was typically $\sim$100 nV in BFO\cite{parsonnet2022nonvolatile} and $\sim$4x larger in lanthanum-substituted BFO\cite{husain2024non}. The $V_{ISHE}$ further goes up by order of magnitude with a large spin-charge conversion efficiency of a metal oxide such as SrIrO$_3$\cite{huang2024manipulating}. This indicates the critical role of the spin-charge conversion efficiency of the SO metal and magnon-spin-transmission channels such as BFO. Here, we propose a heretofore unexplored pathway to exploit both the SO metal and the magnon channel spin transmission efficiency. In the latter, magnons can carry the spins efficiently without loss of information over a long distance by utilizing the dynamics of antiferromagnets order\cite{han2023coherent,lebrun2018tunable} and two-dimensional confinement effects\cite{wei2022giant}. As an efficient approach to two-dimensional magnon confinement in multiferroics, antiferromagnets can provide the additional degree of freedom to control the magnon flow electrically - effects which have been modeled\cite{beairsto2021confined} but are yet to be realized experimentally.
    \begin{figure}[h!]
    	%\centering
\includegraphics[width=1.0\textwidth]{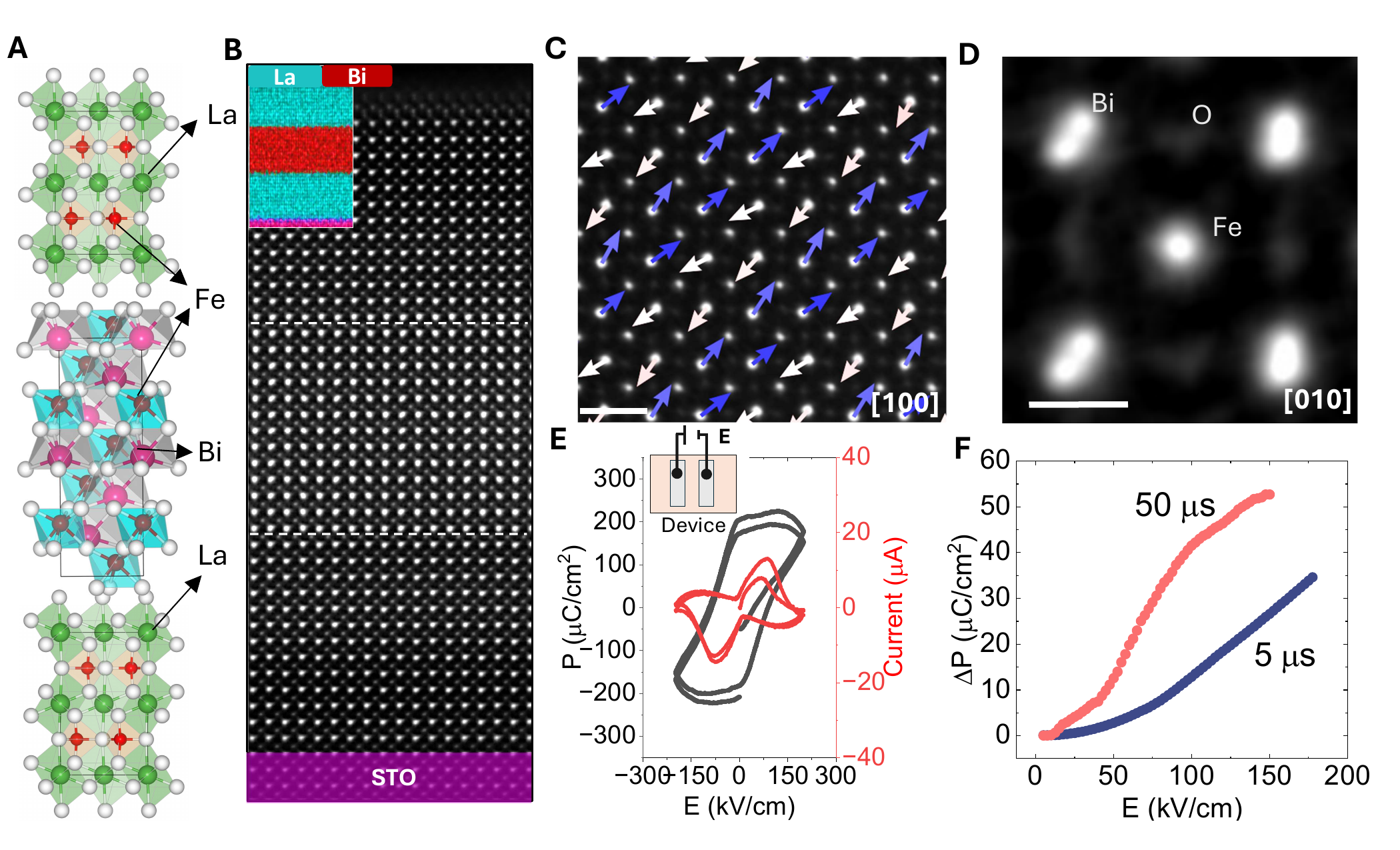}
    	\caption{\textbf{Epitaxial LFO/BFO/LFO heterostructure:} \textbf{(A)} Schematic view of the atomic structures of the LaFeO$_3$ and BiFeO$_3$. \textbf{(B)} HAADF Scanning transmission electron microscopy of the LFO/BFO/LFO heterostructure. Inset shows a false color contrast for elemental mapping measured by energy dispersive X-ray spectroscopy. Elemental diffusion at the interface is appears to be negligible. \textbf{(C)} Phase image of the BFO layer within the heterostructure reconstructed by electron ptychography along the [100] zone axes, scale bar is 5 \AA. The vector map shows the unit cell doubling represents the antipolar phase of BFO. \textbf{(D)} Phase image of the BFO unit cell. Scale bar is 2\AA. \textbf{(E)} Ferroelectric hysteresis measured in LFO/BFO/LFO trilayer sample in lateral geometry (device schematic is in inset). The spacing between the electrodes is 2$\mu$m. Current loops recorded in the same device show the ``butterfly" feature corresponding to the switching  at the coercive field. \textbf{(F)} Switchable polarization measure by PUND (Positive-Up Negative-Down) experiment at 5 and 50$\mu$s voltage pulses.} 
    	\label{fig:1}
    	%\vspace{1000pt}
    \end{figure}

In this work, the fundamental hypothesis is that one can confine magnon transport in the BFO system by sandwiching it between layers of non-polar antiferromagnets such as LaFeO$_3$ (LFO) to induce, in a simplistic picture, confinement of magnon modes in the plane.  This, in turn, would lead to a more efficient transport of spins in the BFO and, consequently, a higher $V_{ISHE}$ at the SO metal could be achieved and controlled by an electric field. We have discovered that this is indeed the case for a model, epitaxial LFO/BFO/LFO trilayer heterostructure, leading to several orders of magnitude enhancement in the $V_{ISHE}$ as compared to a single BFO layer of commensurate thickness. This paper describes these observations and their implications for electric-field-controlled spin-based memory and logic elements\cite{manipatruni2019scalable}.\\

%\textbf{Results}

Cross-sectional high-angle annular dark-field scanning transmission electron microscopy (HAADF-STEM) images (Methods) reveal the atomically precise nature of the STO / LFO (5nm) / BFO (5nm) / LFO (5nm) interfaces (Fig. 1). The inset to this image (Fig.1B) is an atomic-scale resolved energy dispersive X-ray map of the lanthanum and bismuth which authenticates the atomically abrupt interfaces, with negligible inter-diffusion between the layers due to the atomic layer-by-layer growth (fig. S1). Vector mapping reveals that portions of the BFO layer exhibit a distorted BFO (consistent with reciprocal space mapping studies; fig. S2 and S3), reminiscent of the $Pnma$-antiferroelectric phase \cite{mundy2022liberating,caretta2023non}. It has been shown that the interfacial electrostatic boundary conditions (from dielectric layers) imposed on a confined BFO layer, leads to the antipolar phase, which is manifested in the doubling of the unit cell (opposite vector mapping, Fig. 1C and BFO unit cell, Fig.1D). This can be switched into a polar ferroelectric state with an applied electric bias\cite{mundy2022liberating,caretta2023non,borisevich2010suppression} as shown by the ferroelectric hysteresis (Polarization, P electric field, E) (Fig. 1E). Due to the dielectric layer and a thin BFO (5nm), the P-E contains all contributions including the dielectric response and resistive leakage; however, the current-voltage loops show a clear switching around 100 kV/cm. To further validate the ferroelectric switching, PUND (positive-up negative-down) measurements were performed (Fig. 1F)  under short electrical pulses (Methods, fig. S4). The voltage-induced switching at fixed pulse widths (5 $\mu$s and 50 $\mu$s) represents the switchable polarization $\Delta P$, confirming the ferroelectric nature of the BFO(5nm) sandwiched between the LFO layers after electrical switching. Phase-field simulations (fig. S4) support the emergence of polar order from the antipolar ground state in the BFO sandwiched between the LFO layers after the application of an electrical field.

    \begin{figure}[h!]
    	\centering
    	\includegraphics[width=1.0\textwidth]{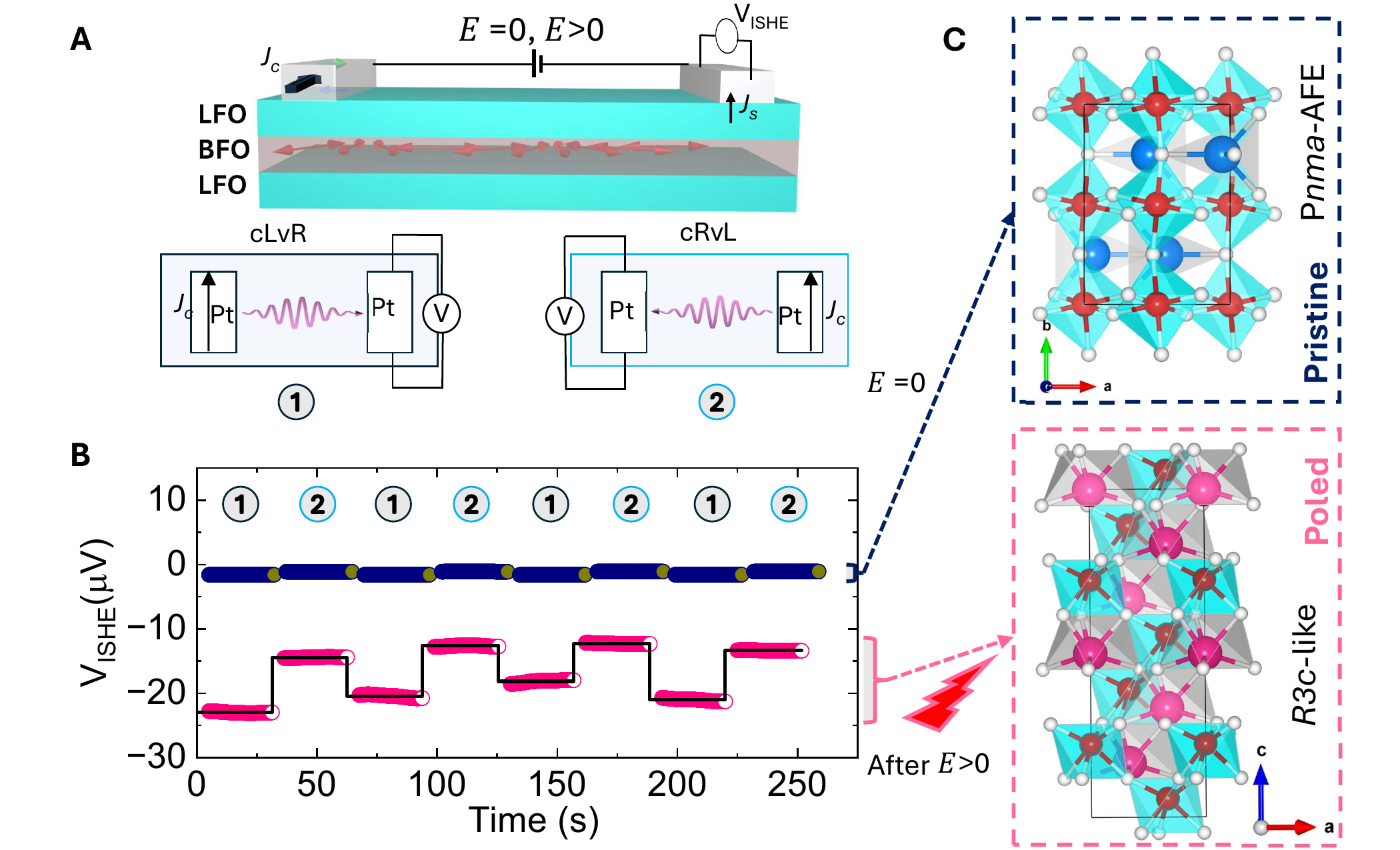}
    	\caption{\textbf{Symmetry breaking and spin transmission:} \textbf{(A)} Schematic of the non-local magnon transport in the LaFeO$_3$/BiFeO$_3$/LaFeO$_3$ with Pt as spin Hall source/detector. $J_c$ and $J_s$, respectively the injected charge current and detected spin current densities. $V_{ISHE}$ is the measured non-local inverse spin Hall voltage. The bottom panel in \textbf{A}, represents the measurement geometry for non-reciprocal spin transport by swapping the source/detector, named as current-left voltage right (cLvR) and current-right voltage-left (cRvL) designated as 1 and 2, respectively. The spin wave drawing represents the right and left propagating magnons in 1 and 2, respectively. \textbf{(B)} Non-local $V_{ISHE}$ measured in the pristine state of the device (when no electric field was applied) and poled state (after electric field applied) enabled magnon transport. The circled number corresponds to the data measured in the cLvR and cRvL geometry. Both data sets in \textbf{(B)} are recorded in the absence of an electric field. Spin texture drawing (top panel) only in the BFO is the representation of the electric field controlled region. LFO is transparent to the electric field response (discussed in text). Lines are guided to the eye. The two measured states (blue and pink) corresponds to the two symmetry states of crystal structure of the BFO shown in \textbf{(C)} corresponding to structural transformation under electrical excitation. Symmetric ($Pnma$-antiferroelectric `AFE') to non-centrosymmetric ($R3c$) like phase emerges after applying electrical pulses.}
    	\label{fig:2}
    	%\vspace{1000pt}
    \end{figure}

Non-local spin transport measurements (first ($\omega$) and second harmonic (2$\omega$)) were designed as illustrated in Fig. 2A (see also Methods and Supplementary text). 
%Metal wires were patterned using spin-Hall metals including platinum (Pt), tungsten (W), and/or SrIrO$_3$ (SIO) to study the effects of spin-charge conversion, both electrically \cite{meisenheimer2024designed} and thermally\cite{husain2024non}. While the electrically injected magnons depend on the spin-to-charge interconversion efficiency of the injector electrode, thermally injected magnons only depend on the temperature gradient generated by Joule heating at the injector. Since the detection process is the same, the signals from the electrically injected magnons depend on both the injector and detector ($i.e.$, direct and inverse charge-to-spin conversion) efficiencies, while for thermally generated magnons the signals are only dependent on the detection (spin-to-charge conversion) efficiency. These two magnon populations are detected as the first ($\omega$) and second harmonic (2$\omega$) responses in the lock-in amplifier. %We mainly focus on the spin-Hall effect-driven charge-to-spin interconversion (Methods) since it avoids the inhomogeneous heat-driven contributions to the signal. 
In these heterostructures, both the antiferromagnetic compounds (LFO and BFO) intrinsically show antiferrodistortive distortions\cite{Nicola2012noncollinear} (responsible for nonpolar oxygen octahedra rotations), while BFO has the additional feature of having a ferroelectric order parameter that gives the electric-field controllability, whereas LFO does not. Importantly, both are canted antiferromagnets due to a large Dzyaloshinskii-Moriya interaction (DMI) which is susceptible to the non-reciprocity in magnon dispersion\cite{cheong2018broken,albrecht2010ferromagnetism}. Additionally, the polar order, which is the highest-energy order parameter in multiferroics, imposes further symmetry constraints\cite{dong2019magnetoelectricity}. As discussed in Fig. 1, the as-grown state is mainly in the antipolar phase, in which the DMI-based canting is minimal due to symmetry considerations\cite{bertaut1968representation}. Thus, before the application of an electric field (in the pristine state), the $V_{ISHE}$ $\simeq$ 100 nV is insignificant as illustrated (Fig.2 and corresponding spin-Seebeck effect data in fig. S7). The two schematics (Current-Left, Voltage-Right `cLvR' $\textcircled{1}$ and Current-Right, Voltage-Left `cRvL' $\textcircled{2}$, Fig. 2A) illustrate the two measurement cases with the source and detector swapped using a switch box. Hereinafter, we call the spin transport process non-reciprocal if the non-local signal in cLvR and cRvL configurations does not have the same magnitude.  

In contrast, after the device is subjected to an electric field of $\sim$300 kV/cm, we observe a strikingly large inverse-spin-Hall response (red data, Fig. 2B). No change in physical parameters such as resistance (of the Pt or BFO) was observed (fig. S6). The output voltage signal has increased to $\simeq$10 $\mu$ V (differential voltage, $\Delta V_{ISHE}$). %The magnon transport gives a negative first harmonic nonlocal signal for this measurement configuration\cite{ZhangNZhang}. 
The data recorded for $\sim$50 sec in each (cLvR/cRvL) configuration with no applied electric field during the measurement is indicative of the non-volatile nature of the magnon propagation, consistent with previous observations\cite{parsonnet2022nonvolatile,husain2024non}. We hypothesize that the significantly larger output voltage (compared to the pristine state) is a consequence of the structural phase transition of the antipolar $Pnma$ phase to a polar $R3c$-like state in BFO after the application of an electric field as depicted in Fig.2C (discussed later in Figure 4 and fig. S7 and S8). We argue that due to the symmetric nature of the antipolar phase in the multiferroic, it should not produce a sizable DMI and therefore, non-reciprocal effects are not expected as evidenced by the very small non-local voltage measured in the pristine state. The large ISHE voltage emerges in the poled state. It is particularly noteworthy that the BFO single layer does not show a measurable spin-Hall effect with Pt\cite{parsonnet2022nonvolatile,husain2024non} and the spin-Seebeck effect for a 100-nm-thick BFO layer is  $\Delta V_{ISHE}$ $\approx$ 100 nV. Thus, the observation of a $\sim$100X larger spin-Hall voltage response of $\sim$10 $\mu$V is exciting. We discuss possible origins for such a ``colossal" enhancement in $V_{ISHE}$ later. 
    \begin{figure}[htbp!]
    	\centering
    	\includegraphics[width=1.0\textwidth]{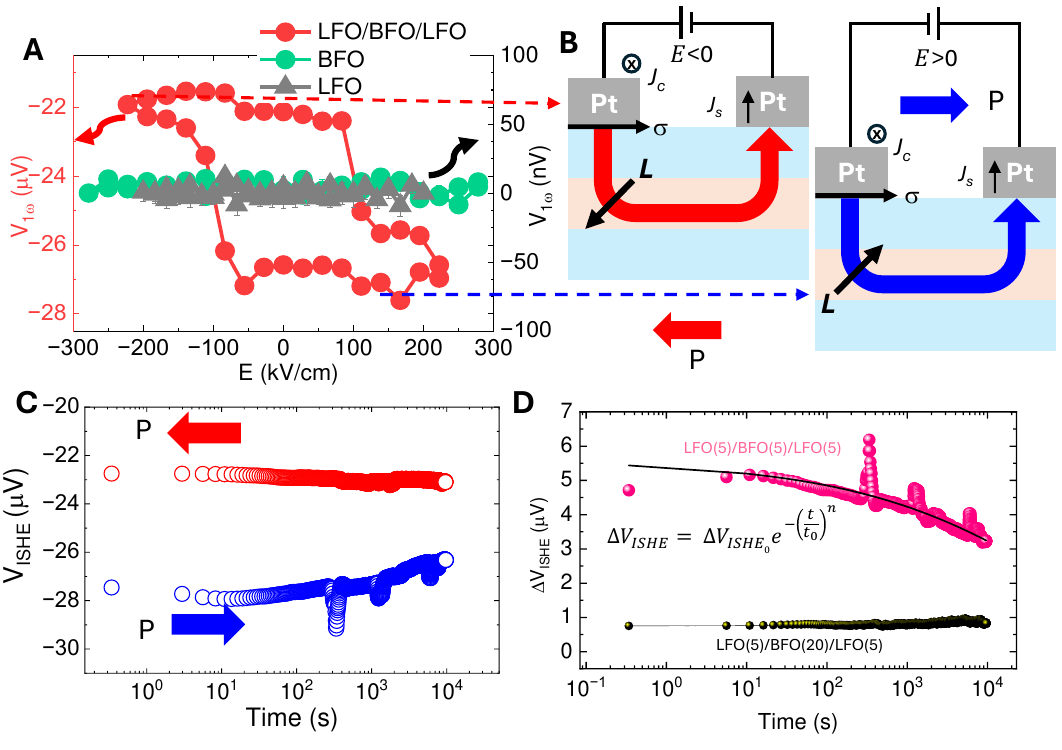}
    	\caption{\textbf{Electric field driven magnon confinement in the trilayer thin films:} \textbf{(A)} Electric field induced magnon driven ISHE voltage in the LFO(5)/BFO(5)/LFO(5) trilayer (red data). V$_{ISHE}$ measured (unshifted) in the BFO and LFO single layers are shown for comparison. \textbf{(B)} Schematics are the representation of the polarization controlled antiferromagnetic vector and corresponding charge-spin conversion mechanism. $L$ represents the Neel vector of the BFO layer. $J_c$ and $J_s$ are the supply charge current and spin Hall generated spin current flowing into the Pt metal. $P$ is the polar order of the magnetoelectric BFO. Red and Blue arrows of the $P$ represent the two polar states of the BFO under opposite electrical pulses as denoted by dotted lines along with the magnon hysteresis. \textbf{(C)} Non-volatile magnon Hall voltage retention measured in the two opposite poled state (using the electric field $\sim$300kV/cm) representing the deterministic switching of the antiferromagnetic and polar order of the BFO and the corresponding magnon transport. \textbf{(D)} The differential of the retention is fit to the stretched exponential fit (equation is the inset). The slow relaxation describes the ferroelectric to antiferroelectric phase transition over a time scale of days. It can switched back to the ferroelectric state repeatedly using an electric field (fig. S7). Data in the bottom in \textbf{(D)} corresponds to the LFO(5nm)/BFO(20nm)/LFO(5nm) does not show any decay within the same time scale.}
 	\label{fig:3}
    	%\vspace{80pt}
    \end{figure}

Armed with this emergent, large voltage signal from the magnon transport, we then proceeded to map out the $V_{ISHE}$ as a function of applied electric field for the LFO/BFO/LFO trilayer heterostructures and compare it with that from a single LFO and BFO layer (and other control samples) (Methods). %The applied field of $\pm$300 kV/cm switches the polar state of the BFO layer (Fig. 1e,f); also demonstrated in prior works\cite{parsonnet2022nonvolatile,husain2024non}; this leads to a change in the state of the antiferromagnetic order and a resulting change in the $V_{ISHE}$. 
This is captured in Fig. 3A. The electric field drives a change in the polar state, and by extension, sets the particular antiferromagnetic state ($L$) (Fig. 3B) due to the strong magnetoelectric coupling\cite{heron2014deterministic}, $\delta (M,L)/\delta E$ determines the magnetic signal driven by the ISHE-conversion process in the Pt (or W,  Supplementary text and fig. S9) or SIO interface. This is enabled by the direction of the electric field controlled polarization. The corresponding evolution of the charge current magnitude dependence (fig. S10) shows a linear change of the magnon output voltage due to the spin-Hall/Seebeck effect. 
The red (blue) U-shaped arrow indicates the low (high) state of the output voltage governed by the $L$ order parameter (Fig. 3B). Since only the BFO responds to the electric field, the hysteretic behavior of the magnons is depicted to be confined within the BFO layer. Supplementary fig. S11 represents the exemplary data measured in about 50 devices from different samples and device fabrication processes. Strikingly, the effects also disappear when the LFO antiferromagnetic layer is replaced by dielectric layers such as SrTiO$_3$ or TbScO$_3$ (fig. S12 a-c). Due to the negligible spin-orbit coupling in copper (Cu), the spin transport is also suppressed when Pt wires are replaced by Cu wires (fig. S12d). Most importantly, the non-local voltage is stable over at least 40 hours (also fig. S7 on a different device) and found to slowly decay (Fig. 3C). The differential voltage ($\Delta V_{ISHE}$) follows a stretched-exponential \cite{kohlrausch1854theorie} (Fig. 3D), $\Delta V_{ISHE} = V_{0} e^{{(-t/t_{0})}^n}$, where $t$ is the time and $t_0$=7.65$\pm$0.01$\times$10$^4$ sec, $V_0$=5.58 $\pm$ 0.01 $\mu$V and the exponent, $n=0.27 $ $\pm$ 0.02, representing the relaxation behavior of the ferroelectric back to the antiferroelectric state\cite{he2007raman,tan2014transformation,ishchuk2000peculiarities}. The relaxation process may vary from hours to days depending on the material system\cite{nadaud2023study,faye2014non,mundy2022liberating,caretta2023non}.  When the BFO thickness in the LFO/BFO/LFO trilayer is increased to 20nm keeping the LFO thickness at 5nm, we observe a primarily polar state in the BFO; correspondingly, a retention time-independent ISHE voltage of $\sim$0.6$\mu$V is observed in the as-grown state (Fig.3D, bottom data), and is only weakly enhanced when poled with an electric field (figs. S13 and S14). It is noteworthy that the spin Hall voltage magnitude is still $\sim$50X higher compared to a single BFO layer of commensurate thickness, indicating the role of magnon confinement, albeit weaker.

    \begin{figure}[htbp!]
    	\centering
    	\includegraphics[width=0.9\textwidth]{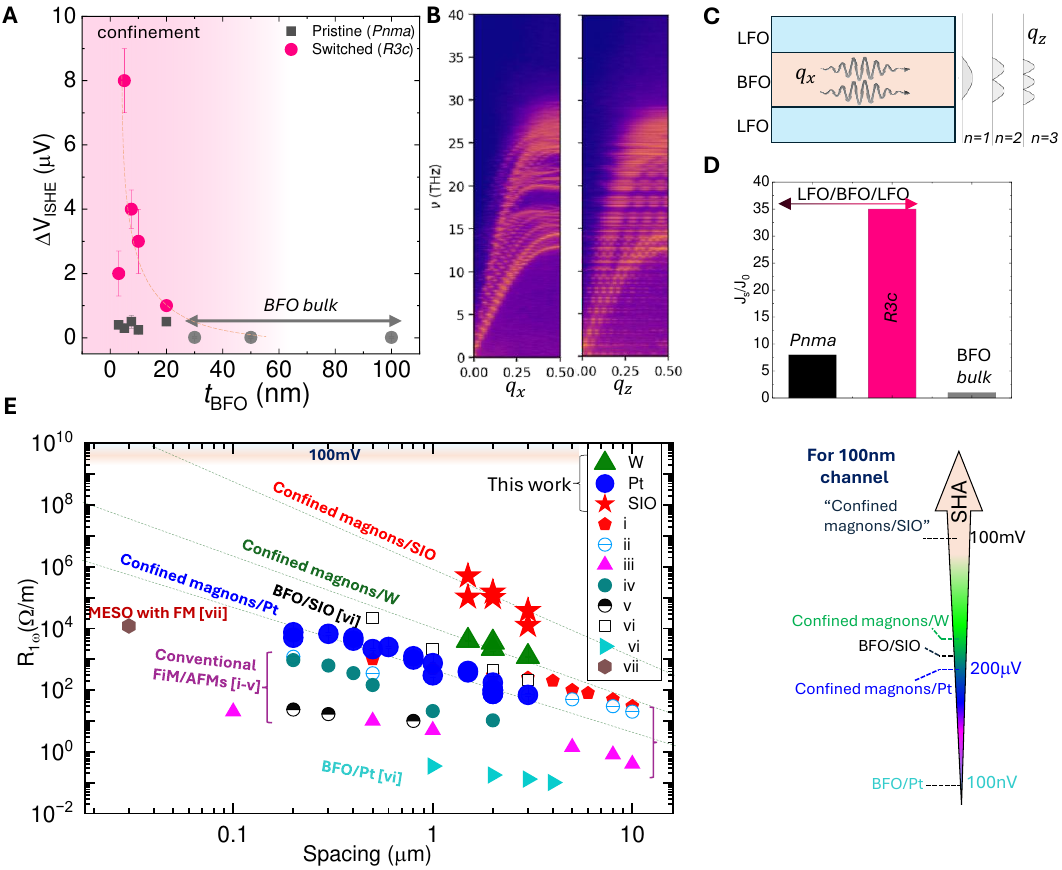}
    	\caption{\textbf{Magnon confinement in BFO and ISHE voltage benchmarking to enable a new MESO:} \textbf{(A)} Non-local ISHE (differential) voltage as a function of BFO thickness (trilayer and the single layer BFO comparison). The data measured in the trilayer is recorded both in the nonpolar and polar states. The dotted line is a guide to the eye. \textbf{(B)} Logarithm of the dynamic spin structure factor $\log(S(q,\nu))$ computed from spin dynamics simulations. High spectral weight $S(q,\nu)$ (orange colors) indicates presence of spin fluctuations at frequency $\nu$ and wave-vector $q$, thereby revealing the dispersion of spin waves. Left panel shows the spectra of spin waves propagating along [100]p.c. with zero out-of-plane momentum ($q_z=0$) . 
        Right panel shows the spectra resolved in $q_z$ for spin waves with zero in-plane momentum ($q_x$=$q_y$=0) showing the coexistence of both propagating and standing wave modes. 
        \textbf{(C)} Schematic representation (left panel) of the magnon confinement in the BFO layer sandwiched between LFO. The interference of scattered magnons results in a standing-wave modulation of the spin precession amplitude throughout the BFO thickness (right panel). The frequency of such excitations propagating along $q_x$ increases with increasing number of standing-wave nodes $n$. \textbf{(D)} Normalized spin current in the trilayer and single layer bulk BFO. \textbf{(E)} Spin Hall resistance calculated from ISHE voltage and scaled with the length of the wire as a function of the SO metal spacing. The benchmarking of the output voltage measured in non-local spin transport in the range of material systems including the ferrimagnets\cite{wei2022giant,cornelissen2015long} `i,ii', hematite\cite{lebrun2018tunable,han2020birefringence} `iii,v', van der Walls antiferromagnet\cite{de2023long} `vi', MESO with a FM\cite{vaz2024voltage} `vii', and multiferroic correlated oxide with SO metals such as Pt, W and SIO\cite{huang2024manipulating} `viii'. Data in blue (Pt), green (W) and red (SIO) belong to this work with confined magnon heterostructure. Lines are the extrapolation of the data for three SO metals. The upright arrow indicates the comparative model of the voltage output that is scaled with the material type. The difference between the BFO/Pt, [BFO/LFO/BFO]/(Pt,W), and BFO/SIO provides the possible pathway to achieve 100 mV with further device lateral or vertical scaling.}
    	\label{fig:4}
    	\vspace{60pt}
    \end{figure}

In the trilayer heterostructures, the LFO layers impart boundary conditions that impact the structural/polar (Fig. 1 and fig. S4) and magnetic response of the BFO\cite{Nicola2012noncollinear,carcan2017phase}. In Fig. 4A, we vary the thickness of BFO keeping the LFO thickness fixed (5 nm for both layers);  the inverse spin Hall voltage asymptotes towards the bulk value \cite{parsonnet2022nonvolatile,husain2024non,huang2024manipulating}. Below $\sim$5 nm, we observe a decrease in the $V_{ISHE}$, which can be attributed to a stronger interfacial scattering.  Such a thickness dependence of the magnon transport points to the essential role of the two-dimensional confinement of magnons in the BFO channel. 

This hypothesis of tunable magnon confinement is supported by effective Hamiltonian-based simulations (Methods, Supplementary text figs. S21, S22, S23). The magnon dispersion obtained from the Landau-Lifshitz-Gilbert spin dynamics simulations\cite{wang2012atomistic} is shown in Fig. 4B. The left (right) panels reveal the spin wave propagating (localized) along $q_x$[100] ($q_z$[001]). Compared to the $R3c$ and $Pnma$ phases of bulk BiFeO$_3$ (fig.~\ref{fig:magnonsBulk} and Supplementary text), both trilayers feature a slightly reduced magnon gap at the zone center. The magnons with non-zero out-of-plane momentum are scattered at the LFO/BFO interface experiencing an almost total internal reflection. In turn, the interference of incident and reflected spin excitations gives rise to standing wave-like magnons (Fig. 4C) in the z-direction. While the confined magnons do not transfer energy in the out-of-plane direction, they do propagate in-plane as evidenced by the distinguishable dispersion curves shown in Fig. 4B. For the polar BFO case, the in-plane group velocity of such excitations remains high, thereby providing additional spin current channels. We also find that the trilayer with switched BFO features a higher spin pumping current at the top electrode/LFO interface. The efficiency of the dc spin current $J_S$ strongly depends on the magnon polarization and in the case of antiferromagnets, is proportional to the sum of the Néel vector $L$ and magnetization $M$ contributions $J_S \sim M \times dM/dt + L \times dL/dt$ ~\cite{cheng2014}. The estimated efficiency of spin pumping for the cases of polar and non-polar trilayers as well as the $R3c$ phase of bulk BFO (Fig. 4D) shows a significant enhancement of spin pumping in the case of trilayer system with the polar BFO. We attribute such enhancement to the long-wavelength confined magnons that exhibit high amplitude of spin and magnetization precession at the surface of the top LFO layer (fig. S17c). In addition, it can also be shown theoretically that a model of spin transport based on both interface-confined and BFO-confined magnons supports a dramatic enhancement of spin conductivity as the BFO layer thickness $t_{BFO}$ decreases (Supplementary text, fig. s25, s26). As  $t_{BFO}$ decreases, the BFO-confined subbands are depopulated, leading to reduction in magnon-magnon scattering and spin current $\langle J_S\rangle \propto \left(\frac{\lambda_T}{t_{BFO}}\right)^{2}$, with $\lambda_T$ the thermal wavelength for antiferromagnetic magnons (Eq. S14). This provides a qualitative explanation of the increase of spin current observed in Fig. 4A.

\noindent\textbf{Discussion and Outlook}\\
We have presented a pathway by which there is a $\sim$100X enhancement in the spin transmission, leading to a corresponding enhancement in the $V_{ISHE}$, and its manipulation by an electric field in a perfectly epitaxial LFO/BFO/LFO all antiferromagnet heterostructure. The stark differences between a single BFO layer and the trilayer heterostructure (Fig. 3A, and Fig. 4A) point to the key role of the LFO layers in confining the magnons within the BFO substantiated with the fundamental understanding of the origin of such an enhancement employing the model Hamiltonian. A comparative benchmarking of the spin-Hall data reported in conventional ferrimagnetic insulators, hematite, orthoferrites, and correlated oxides is presented (Fig. 4D). The voltage is normalized with the supply current ($I_{ac}$) and the length of the wire ($L$) as $R_{1\omega}=V_{1\omega}/(I_{ac}\times L)$ for comparison. The overall magnitude of the output voltage has been shown to depend on the spacing between the spin orbit metal wires and the spin-Hall angle (for platinum, tungsten and SrIrO$_3$). An order of magnitude enhancement has been observed going from platinum $\rightarrow$ SrIrO$_3$ (from blue circle  to red star symbols), due to the difference in the intrinsic spin-Hall angle between Pt and SrIrO$_3$. From a practical perspective, this discovery, in conjunction with the significant enhancements ($\sim$10-30X) in the spin-to-charge conversion using epitaxial complex oxides\cite{huang2024manipulating} opens up a wide spectrum of opportunities for both fundamental science of magnon confinement in such epitaxial heterostructures as well as enhancing the $V_{ISHE}$ towards 100 mV\cite{manipatruni2019scalable} to enable low-voltage, logic-in memory functionalities.\\

%    \noindent\textbf{References}
    \bibliographystyle{sciencemag}
    \bibliography{manuscript}

\begin{thebibliography}{10}
\providecommand{\url}[1]{\texttt{#1}}
\expandafter\ifx\csname urlstyle\endcsname\relax
  \providecommand{\doi}[1]{doi:\discretionary{}{}{}#1}\else
  \providecommand{\doi}{doi:\discretionary{}{}{}\begingroup \urlstyle{rm}\Url}\fi

\bibitem{baltz2018antiferromagnetic}
V.~Baltz, \emph{et~al.}, Antiferromagnetic spintronics. \emph{Rev. Mod. Phys.} \textbf{90}~(1), 015005 (2018).

\bibitem{jungwirth2016antiferromagnetic}
T.~Jungwirth, X.~Marti, P.~Wadley, J.~Wunderlich, Antiferromagnetic spintronics. \emph{Nature nanotechnology} \textbf{11}~(3), 231--241 (2016).

\bibitem{nvemec2018antiferromagnetic}
P.~N{\v{e}}mec, M.~Fiebig, T.~Kampfrath, A.~V. Kimel, Antiferromagnetic opto-spintronics. \emph{Nat. Phys.} \textbf{14}~(3), 229--241 (2018).

\bibitem{wadley2016electrical}
P.~Wadley, \emph{et~al.}, Electrical switching of an antiferromagnet. \emph{Science} \textbf{351}~(6273), 587--590 (2016).

\bibitem{vzelezny2014relativistic}
J.~{\v{Z}}elezn{\`y}, \emph{et~al.}, Relativistic \textrm{N}{\'e}el-order fields induced by electrical current in antiferromagnets. \emph{Phys. Rev. Lett.} \textbf{113}~(15), 157201 (2014).

\bibitem{bhattacharjee2014prediction}
S.~Bhattacharjee, S.~Singh, D.~Wang, M.~Viret, L.~Bellaiche, Prediction of novel interface-driven spintronic effects. \emph{Journal of Physics: Condensed Matter} \textbf{26}~(31), 315008 (2014).

\bibitem{duttagupta2020spin}
S.~DuttaGupta, \emph{et~al.}, Spin-orbit torque switching of an antiferromagnetic metallic heterostructure. \emph{Nat. Comm.} \textbf{11}~(1), 5715 (2020).

\bibitem{lebrun2018tunable}
R.~Lebrun, \emph{et~al.}, Tunable long-distance spin transport in a crystalline antiferromagnetic iron oxide. \emph{Nature} \textbf{561}~(7722), 222--225 (2018).

\bibitem{wei2022giant}
X.-Y. Wei, \emph{et~al.}, Giant magnon spin conductivity in ultrathin yttrium iron garnet films. \emph{Nat. Mat.} \textbf{21}~(12), 1352--1356 (2022).

\bibitem{han2023coherent}
J.~Han, R.~Cheng, L.~Liu, H.~Ohno, S.~Fukami, Coherent antiferromagnetic spintronics. \emph{Nat. Mat.} \textbf{22}~(6), 684--695 (2023).

\bibitem{parsonnet2022nonvolatile}
E.~Parsonnet, \emph{et~al.}, Nonvolatile electric field control of thermal magnons in the absence of an applied magnetic field. \emph{Phys. Rev. Lett.} \textbf{129}~(8), 087601 (2022).

\bibitem{huang2024manipulating}
X.~Huang, \emph{et~al.}, Manipulating chiral spin transport with ferroelectric polarization. \emph{Nat. Mat.} p. 898–904 (2024).

\bibitem{hamdi2023spin}
M.~Hamdi, F.~Posva, D.~Grundler, Spin wave dispersion of ultra-low damping hematite ($\alpha-\textrm{F}e_{2}\textrm{O}_{3}$) at \textrm{GH}z frequencies. \emph{Phys. Rev. Mat.} \textbf{7}~(5), 054407 (2023).

\bibitem{AlbertRMP}
A.~Fert, R.~Ramesh, V.~Garcia, F.~Casanova, M.~Bibes, Electrical control of magnetism by electric field and current-induced torques. \emph{Rev. Mod. Phys.} \textbf{96}, 015005 (2024).

\bibitem{salahuddin2018era}
S.~Salahuddin, K.~Ni, S.~Datta, The era of hyper-scaling in electronics. \emph{Nat. Elect.} \textbf{1}~(8), 442--450 (2018).

\bibitem{manipatruni2019scalable}
S.~Manipatruni, \emph{et~al.}, Scalable energy-efficient magnetoelectric spin--orbit logic. \emph{Nature} \textbf{565}~(7737), 35--42 (2019).

\bibitem{vaz2024voltage}
D.~C. Vaz, \emph{et~al.}, Voltage-based magnetization switching and reading in magnetoelectric spin-orbit nanodevices. \emph{Nat. Comm.} \textbf{15}~(1), 1902 (2024).

\bibitem{chai2024voltage}
Y.~Chai, \emph{et~al.}, Voltage control of multiferroic magnon torque for reconfigurable logic-in-memory. \emph{Nat. Comm.} \textbf{15}~(1), 5975 (2024).

\bibitem{cherifi2014electric}
R.~Cherifi, \emph{et~al.}, Electric-field control of magnetic order above room temperature. \emph{Nat. Mat.} \textbf{13}~(4), 345--351 (2014).

\bibitem{chu2008electric}
Y.-H. Chu, \emph{et~al.}, Electric-field control of local ferromagnetism using a magnetoelectric multiferroic. \emph{Nat. Mat.} \textbf{7}~(6), 478--482 (2008).

\bibitem{husain2024non}
S.~Husain, \emph{et~al.}, Non-volatile magnon transport in a single domain multiferroic. \emph{Nat. Comm.} \textbf{15}~(1), 5966 (2024).

\bibitem{beairsto2021confined}
S.~Beairsto, M.~Cazayous, R.~S. Fishman, R.~de~Sousa, Confined magnons. \emph{Phys. Rev. B} \textbf{104}~(13), 134415 (2021).

\bibitem{mundy2022liberating}
J.~A. Mundy, \emph{et~al.}, Liberating a hidden antiferroelectric phase with interfacial electrostatic engineering. \emph{Science Advances} \textbf{8}~(5), eabg5860 (2022).

\bibitem{caretta2023non}
L.~Caretta, \emph{et~al.}, Non-volatile electric-field control of inversion symmetry. \emph{Nat. Mat.} \textbf{22}~(2), 207--215 (2023).

\bibitem{borisevich2010suppression}
A.~Y. Borisevich, \emph{et~al.}, Suppression of Octahedral Tilts and Associated Changes in Electronic Properties at Epitaxial Oxide Heterostructure Interfaces. \emph{Phys. Rev. Lett.} \textbf{105}, 087204 (2010).

\bibitem{Nicola2012noncollinear}
C.~Weingart, N.~Spaldin, E.~Bousquet, Noncollinear magnetism and single-ion anisotropy in multiferroic perovskites. \emph{Phys. Rev. B} \textbf{86}~(9), 094413 (2012).

\bibitem{cheong2018broken}
S.-W. Cheong, D.~Talbayev, V.~Kiryukhin, A.~Saxena, Broken symmetries, non-reciprocity, and multiferroicity. \emph{npj Quantum Mat.} \textbf{3}~(1), 19 (2018).

\bibitem{albrecht2010ferromagnetism}
D.~Albrecht, \emph{et~al.}, Ferromagnetism in multiferroic $\textrm{B}i\textrm{F}e\textrm{O}_{3}$ films: a first-principles-based study. \emph{Physical Review B—Condensed Matter and Materials Physics} \textbf{81}~(14), 140401 (2010).

\bibitem{dong2019magnetoelectricity}
S.~Dong, H.~Xiang, E.~Dagotto, Magnetoelectricity in multiferroics: a theoretical perspective. \emph{Nat. Sci. Rev.} \textbf{6}~(4), 629--641 (2019).

\bibitem{bertaut1968representation}
E.~Bertaut, Representation analysis of magnetic structures. \emph{Acta Crystallographica Section A: Crystal Physics, Diffraction, Theoretical and General Crystallography} \textbf{24}~(1), 217--231 (1968).

\bibitem{heron2014deterministic}
J.~Heron, \emph{et~al.}, Deterministic switching of ferromagnetism at room temperature using an electric field. \emph{Nature} \textbf{516}~(7531), 370--373 (2014).

\bibitem{kohlrausch1854theorie}
R.~Kohlrausch, Theorie des elektrischen R{\"u}ckstandes in der Leidener Flasche. \emph{Annalen der Physik} \textbf{167}~(2), 179--214 (1854).

\bibitem{he2007raman}
H.~He, X.~Tan, Raman spectroscopy study of the phase transitions in \textrm{P}b$_{0.99}$\textrm{N}b$_{0.02}$[(\textrm{Z}r$_{0.57}$\textrm{S}n$_{0.43}$)$_{1-y}$\textrm{T}i$_y$]$_{0.98}$\textrm{O}$_3$ ceramics. \emph{Journal of Physics: Condensed Matter} \textbf{19}~(13), 136003 (2007).

\bibitem{tan2014transformation}
X.~Tan, \emph{et~al.}, Transformation toughening in an antiferroelectric ceramic. \emph{Acta materialia} \textbf{62}, 114--121 (2014).

\bibitem{ishchuk2000peculiarities}
V.~Ishchuk, O.~Belichenko, O.~Nikolov, V.~Sobolev, Peculiarities of ferro-antiferroelectric phase transitions. 6. \textrm{E}xperiments on low-frequency dynamics of interphase domain walls. \emph{Ferroelectrics} \textbf{248}~(1), 107--122 (2000).

\bibitem{nadaud2023study}
K.~Nadaud, \emph{et~al.}, Study of the long time relaxation of the weak ferroelectricity in $\textrm{P}b\textrm{Z}r\textrm{O}_3$. \emph{Thin Solid Films} \textbf{773}, 139817 (2023).

\bibitem{faye2014non}
R.~Faye, H.~Liu, J.-M. Kiat, B.~Dkhil, P.-E. Janolin, Non-ergodicity and polar features of the transitional phase in lead zirconate. \emph{Applied Physics Letters} \textbf{105}~(16) (2014).

\bibitem{cornelissen2015long}
L.~Cornelissen, J.~Liu, R.~Duine, J.~B. Youssef, B.~Van~Wees, Long-distance transport of magnon spin information in a magnetic insulator at room temperature. \emph{Nat. Phys.} \textbf{11}~(12), 1022--1026 (2015).

\bibitem{han2020birefringence}
J.~Han, \emph{et~al.}, Birefringence-like spin transport via linearly polarized antiferromagnetic magnons. \emph{Nat. Nanotech.} \textbf{15}~(7), 563--568 (2020).

\bibitem{de2023long}
D.~K. De~Wal, \emph{et~al.}, Long-distance magnon transport in the van der Waals antiferromagnet $\textrm{C}r\textrm{PS}_{4}$. \emph{Phy. Rev. B} \textbf{107}~(18), L180403 (2023).

\bibitem{carcan2017phase}
B.~Carcan, \emph{et~al.}, Phase diagram of $\textrm{B}i\textrm{F}e\textrm{O}_{3}/\textrm{L}a\textrm{F}e\textrm{O}_{3}$ superlattices: antiferroelectric-like state stability arising from strain effects and symmetry Mismatch at Heterointerfaces. \emph{Adv. Mat. Int.} \textbf{4}~(11), 1601036 (2017).

\bibitem{wang2012atomistic}
D.~Wang, J.~Weerasinghe, L.~Bellaiche, Atomistic molecular dynamic simulations of multiferroics. \emph{Phys. Rev. Lett.} \textbf{109}~(6), 067203 (2012).

\bibitem{cheng2014}
R.~Cheng, J.~Xiao, Q.~Niu, A.~Brataas, Spin Pumping and Spin-Transfer Torques in Antiferromagnets. \emph{Phys. Rev. Lett.} \textbf{113}, 057601 (2014).

\end{thebibliography}

\noindent\textbf{Acknowledgments}\\
S. H., X.L., R.R., and L.W.M acknowledge that this research was sponsored by the Army Research Laboratory and was accomplished under Cooperative Agreement Number W911NF-24-2-0100. The views and conclusions contained in this document are those of the authors and should not be interpreted as representing the official policies, either expressed or implied, of the Army Research Laboratory or the U.S. Government. The U.S. Government is authorized to reproduce and distribute reprints for Government purposes notwithstanding any copyright notation herein. L.W.M, M.R., A.R. L.Q.C. and D.G.S. are supported by the Army Research Office under the ETHOS MURI via cooperative agreement W911NF-21-2-0162. We acknowledge partial support from the U.S. Department of Energy, Office of Science, Office of Basic Energy Sciences, Materials Science and Engineering Division and the U.S. Department of Energy, Office of Science, Advanced Scientific Computing Research (ASCR) program under Contract No. DE-AC02-05-CH11231. L.C. acknowledges support from the National Science Foundation under Grant No. OIA-2327352. X. L. acknowledges support from the Rice Advanced Materials Institute (RAMI) at Rice University as a RAMI Postdoctoral Fellow. S.O., L.W.M., Y.H., and R.R. acknowledge support from the National Science Foundation via Grant DMR-2329111. Y.H. acknowledges the support of the Welch Foundation (C-2065). This research used resources of the Advanced Light Source, which is a DOE Office of Science User Facility under contract No. DE-AC02-05CH11231. S.K.K. and H.P. are supported by the Brain Pool Plus Program through the National Research Foundation of Korea funded by the Ministry of Science and ICT (2020H1D3A2A03099291). R.d.S. acknowledges financial support from NSERC (Canada) through its Discovery Program (Grant No. RGPIN-2020-04328).  
S.P., Y.N. and L.B. also thank the Vannevar Bush Faculty Fellowship (VBFF) Grant No. N00014-20-1–2834 from the Department of Defense, the MonArk NSF Quantum Foundry supported by the National Science Foundation Q-AMASE-i Program under NSF Award No. DMR-1906383, the MURI
ETHOS Grant No. W911NF-21-2-0162 from the Army Research Office (ARO), and technical support from the Arkansas High Performance Computing Center (AHPCC). S.P., Y.N, and L.B. also gratefully acknowledge stimulating discussions with Prof. Bin Xu and Dr. Menghui Xia on model Hamiltonian calculations.
B.Z. and H.W. were supported by the U.S. Department of Energy (DOE), Office of Science (SC), Basic Energy Sciences, Materials Sciences and Engineering Division in the Advanced Light Source, a DOE-SC User Facility operated by Argonne National Laboratory under contract No. DE-AC02-06CH11357. We thank Yi Jiang (Argonne National Laboratory) for helping to set-up the Ptychography imaging. We thank Akash Surampalli (Rice) and Yogesh Kumar (UC Berkeley) for helping with controlled sample characterizations. We acknowledge fruitful ongoing discussions with Professors Nicola Spaldin (ETH), Yaroslav Tserkovnyak (UCLA), Gregory A. Fiete (Northeastern), Sergey Prosandeev (Arkansas) and Dr. Wesley Roberts (Northeastern). \\
\\
    \noindent\textbf{Author contributions}\\
    S.H., M.R., D.S., and R.R. conceived the idea. M. R. synthesized the samples and analyzed the data supervised by D.S. S.H. designed and performed the metal synthesis, device fabrication, and measurements. X.L. performed the electron microscopy and Ptychography supervised by Y.H. S.K.O. helped in the microscopy. A.R. and L.Q.C. performed the phase field simulations. B.Z. and H.W. performed synchrotron X-ray diffraction measurements. K.D. helped in the e-beam lithography of controlled samples supervised by S.S. S.P. and Y.N. computed and analyzed ab initio based magnon spectra and discussed the results with R.d.S and L.B. H.W.P, S.K.K., S.H., Z.Y. and R.d.S. developed an interpretation of the data based on models of confined magnons. S.H. and M.R. wrote the manuscript. M.B. helped with data analysis, interpretation, and manuscript preparation. R.R. ran the whole project.

    \noindent\textbf{Competing interests}:  There are no competing interests to declare.
    
    \noindent\textbf{Data availability}:   The data that support the findings of this study are available from the corresponding authors upon reasonable request.\\  
    \noindent\textbf{Additional information}:    Correspondence and requests for materials should be addressed to S.H. and R.R.\\

\subsection*{Supplementary materials}
Materials and Methods\\
Supplementary Text\\
Figs. S1-S26 \\
Tables S1\\
%References \textit{(43-\arabic{enumiv})}

\end{document}